\documentclass{elsart}
\usepackage{cite}      
\usepackage{graphicx}  
\usepackage{subfigure} 
\usepackage{amsmath}   
\usepackage{amssymb}
\usepackage{array}

\begin{document}

\renewcommand{\labelenumi}{\arabic{enumi}}
\renewcommand{\labelitemi}{-}

\begin{frontmatter}



\title{Pixel Detectors for Charged Particles}
\thanks[1]{Bonn University, Physikalisches Institut,
Nussallee 12, D-53115 Bonn, Germany}

\author{N.~Wermes}
\address{Bonn University, Physikalisches Institut}
\thanks[2]{Work supported by the German Ministerium f{\"u}r Bildung,
              Wissenschaft, Forschung und Technologie (BMBF) under contract
              no.~06 HA6PD1.
       }
\ead{wermes@uni-bonn.de}
\ead[url]{http:hep1.physik.uni-bonn.de}

\begin{abstract}
Pixel Detectors, as the current technology of choice for the innermost
vertex detection, have reached a stage at which large detectors have
been built for the LHC experiments and a new era of developments,
both for hybrid and for monolithic or semi-monolithic pixel
detectors is in full swing. This is largely driven by the
requirements of the upgrade programme for the superLHC and by other collider experiments which plan
to use monolithic pixel detectors for the first time. A review on current pixel detector
developments for particle tracking and vertexing is given, comprising hybrid pixel detectors for superLHC with its own challenges in radiation and rate,
as well as on monolithic, so-called active pixel detectors, including MAPS
and DEPFET pixels for RHIC and superBelle.
\end{abstract}

\end{frontmatter}

\section{Introduction}
\begin{figure}[thb]
\begin{center}
\subfigure[Hybrid pixel]{
    \includegraphics[width=0.35\textwidth]{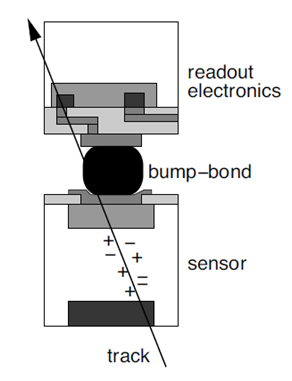}
    \label{sketches_a}}
    \hskip 1cm
\subfigure[Monolithic active pixel]{
    \includegraphics[width=0.5\textwidth]{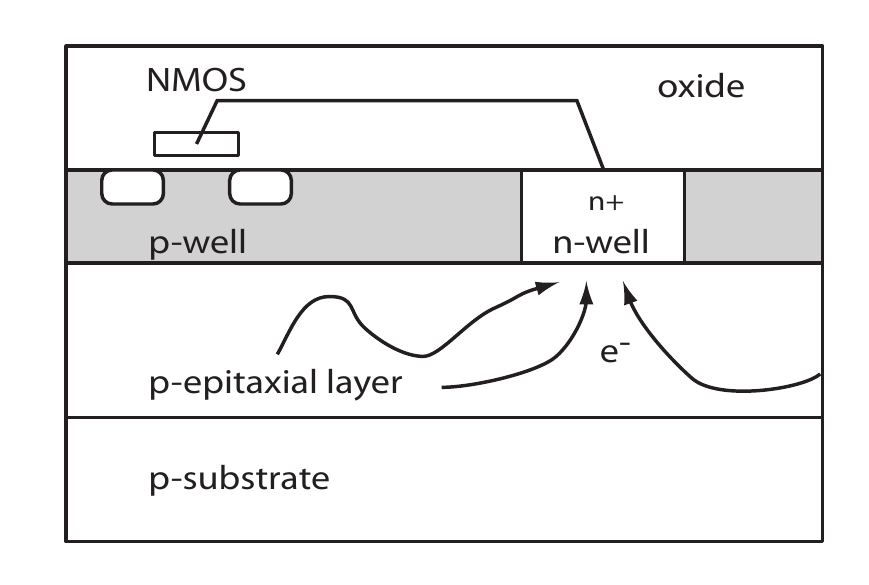}
    \label{sketches_b}}
\end{center}
\caption[]{\label{sketches} Classification of current pixel technologies: (a) Hybrid Pixels which mate sensor and electronics IC by conducting bumps, (b) Monolithic Pixels in which electronics and senor are one entity.}
\end{figure}

One can subdivide the developments of pixel detectors for charged particle detection into two categories: (a) hybrid pixel detectors, in which the particle sensitive volume, the sensor, and the readout IC are separate entities, and (b) monolithic or semi-monolithic detectors in which these are parts of one monolithic block. Both types are illustrated in fig.~\ref{sketches}.
All large detectors built so far, in particular those of the LHC experiments~\cite{ATLAS-pixel-paper_2008,CMS-pixel-paper_2008,ALICE-pixel-paper_2008}, are of the hybrid pixel type. The charge signal, generated in the fully depleted bulk of the sensor, is amplified, filtered and temporarily stored in the pixel cell, until it is transferred at the end of a column of the column-wise organized R/O-chip. Here it waits for the trigger to arrive. Hybrid pixel detectors are now a matured  technology. They exhibit high rate capabilities and have been shown to be radiation tolerant to fluence levels of 10$^{15}$ particles per cm$^2$.
Characteristics of the LHC hybrid pixel detectors which would mostly benefit from improvement are: their moderate spatial resolution ($\sim$10$\mu$m) and their relatively large material budget. The radiation length X$_0$ per layer of the large ($>$ 1m$^2$) ATLAS and CMS pixel detectors is in the order of 3\% X$_0$, not the least due to their overall size and the fact that they must be are operated at -6~$^\circ$C, while for ALICE the material budget is more than a factor of 2 less. The ALICE detector is much smaller (0.2 m$^2$) and can be operated at room temperature.
Hybrid pixel developments for super-LHC (sLHC) detectors are discussed in chapter~\ref{hybrid-pixels}.

Monolithic pixel detector designs, on the other hand, usually operate in frame readout mode. The generated charge is received at a collecting diode connecting to the gate of a transistor. The row of the pixel matrix is selected and the selected pixels are then read out column-wise. After readout the row is reset. We distinguish monolithic CMOS active pixels in which the same substrate is used for charge collection and steering/readout electronics, and DEPFET active pixels which contain a single transistor in a fully depleted bulk, but steering and R/O is done separately on external chips. On first sight the monolithic approach is very appealing. The realization, however, especially the ultimate goal -- the combination of charge collection in a fully depleted bulk and full CMOS electronics in the same substrate -- is not yet reached, although promising developments do exist. The various monolithic developments are discussed in chapter~\ref{monolithic-pixels}.

Table~\ref{rates} compares particle rates and fluences for vertex detector environments at the most relevant running and planned collider experiments.
\begin{table}[!htb]
\begin{center}
\label{rates}
\begin{tabular}[htb]{cccccc}
\hline
& luminosity & BX time & rate & fluence & ion. dose \\
& (cm$^{-2}$ s$^{-1}$) & (ns) & (kHz/mm$^2$) & (n$_{eq}$/cm$^2$) & (kGy) \\
\hline
LHC & 10$^{34}$ & 25 & 1000 & 10$^{15}$ & 790~\cite{ATLAS-pixel-paper_2008} \\
superLHC & 10$^{36}$ & 25 & 10000 & 10$^{16}$ & 5000 \\
superBelle & 10$^{35}$ & 2 & 400 & $\sim$3 $\times$10$^{12}$ & 50 \\
ILC & 10$^{34}$ & 350 & 250 & 10$^{12}$ & 4 \\
STAR@RHIC & 8$\times$10$^{27}$ & 110 & 3.8 & 5 $\times$ 10$^{13}$ & 15 \\
\hline \\
\end{tabular}
\caption{Particle rates and fluences for various collider experiments at the position of the innermost pixel layer. BX is the abbreviation for bunch crossing. The fluences and doses are given for the full life times, i.e. 7 years of design luminosity for LHC and sLHC, 5 years for RHIC and superBelle, and 10 years for the ILC.}
\end{center}
\end{table}
The very high rate and fluence values at LHC and sLHC render hybrid pixels the only choice for further R\&D development. For new vertex detector developments at RHIC and superBelle for which the radiation environment is less fierce and the rates are lower, monolithic pixel detectors have come to the level of concrete proposals.
\section{Hybrid pixel detector development for superLHC}\label{hybrid-pixels}
At the superLHC~\cite{sLHC}, hybrid pixel detectors must be used in the innermost detector volume up to radial distances from the interaction point of about 30 cm.
The challenges to be faced are: sensors which stand radiation levels up to 10$^{16}$ n$_{eq}$/cm$^2$, chip architectures facing output data rates of 320 MHz, and modules and support structures with thicknesses in the order of 1.5$\%$ -- 2$\%$ X$_0$, rather than the current figures of $>$3$\%$.
Studies of the radiation tolerance of pixel sensors has identified three possible solutions for the sLHC fluences: so called 3D silicon sensors, CVD diamond sensors, and planar n$^+$ in p- or n-bulk silicon sensors.

\subsubsection*{3D silicon pixels}
\begin{figure}[thb]
\begin{center}
\subfigure[3D sensor arrangement for ATLAS pixel geometry]{
    \includegraphics[width=0.45\textwidth]{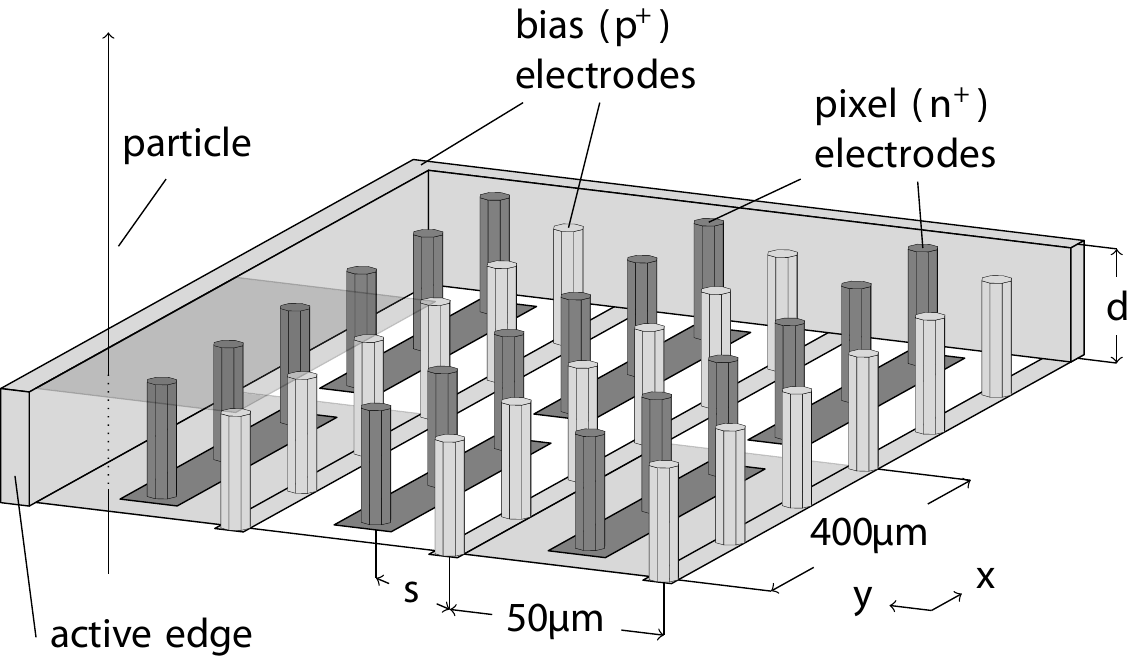}
    \label{3D-pixels-a}}
    \hskip 1cm
    \subfigure[A typical 3D structure]{
    \includegraphics[width=0.4\textwidth]{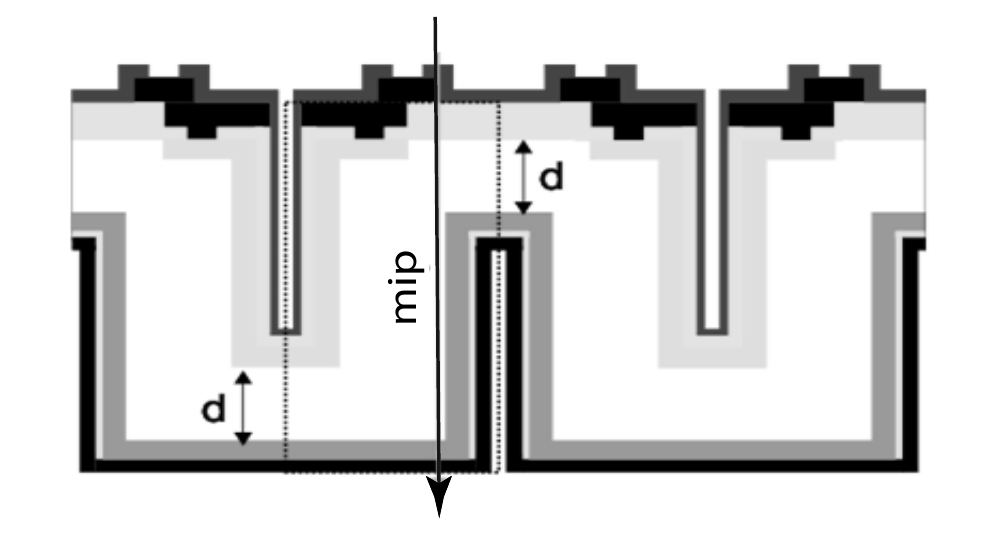}
    \label{3D-pixels-b}}
\end{center}
\caption[]{\label{3D-pixels} (a) 3D pixel layout~\cite{mathes-3D_2008} which fits the ATLAS 50$\times$400 $\mu$m$^2$ pixel pattern of the ATLAS pixel chip (note the different x,y scales), (b) a typical 3D structure with vertical column electrodes (CNM Barcelona).}
\end{figure}

In the 3D silicon sensor concept~\cite{3D_1997} the electrode implants are vertically placed as columns in a high ohmic Si substrate.
Fig.~\ref{3D-pixels}(a) shows a layout footprint which has been mated by solder bump bonding~\cite{IZM1} to the ATLAS pixel chip~\cite{FEI3}. It has three column electrodes per 50 $\mu$m $\times$ 400 $\mu$m pixel cell. Several layout variations are currently being investigated at different vendors, also those in which the electrodes are processed from both sides, not completely reaching through the substrate as shown in fig.~\ref{3D-pixels}(b).

\begin{figure}[thb]
\begin{center}
\includegraphics[width=0.9\textwidth]{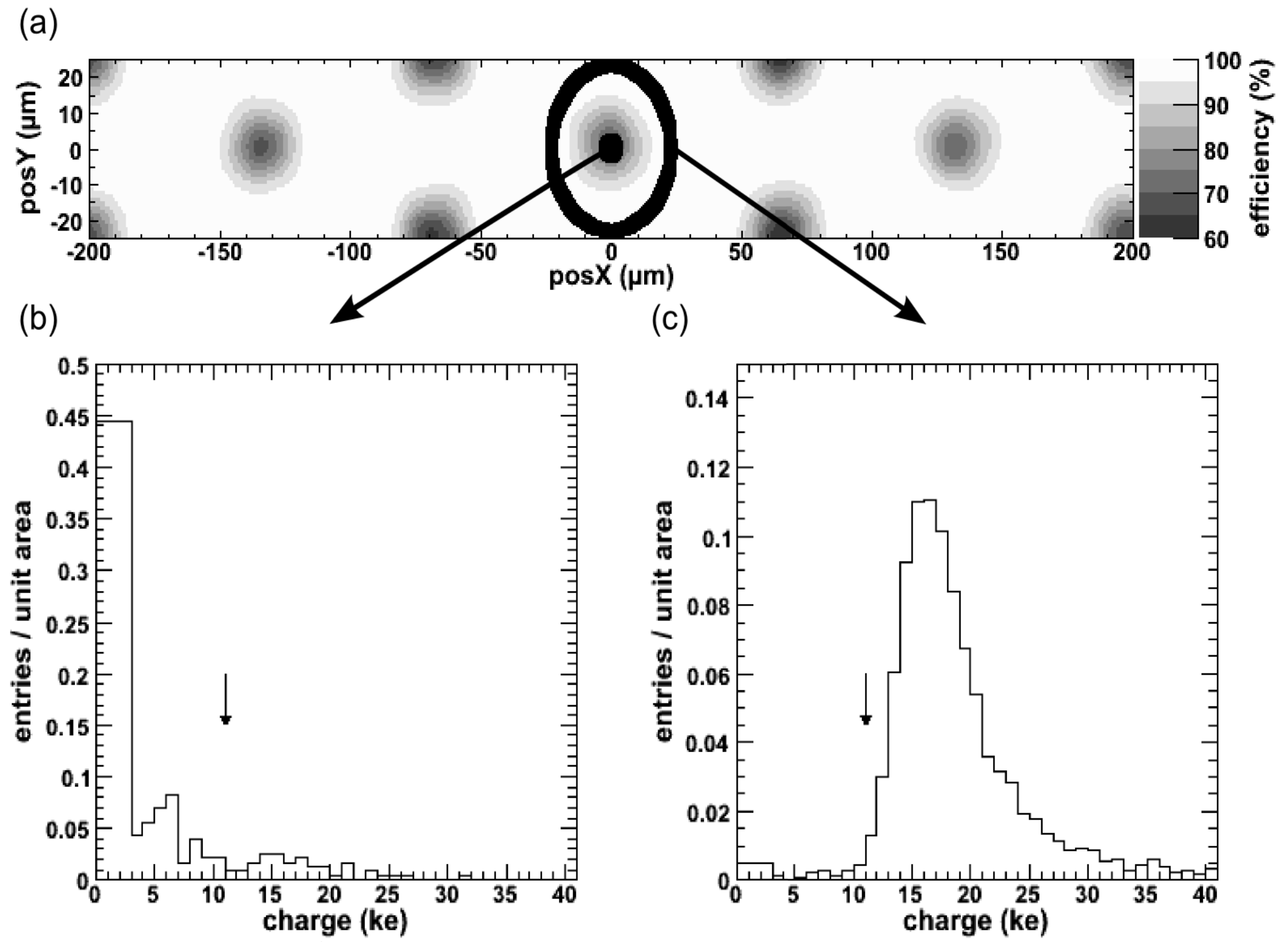}
\end{center}
\caption[]{\label{3D-testbeam} (a) Efficiency map of a 3D pixel sensor studied in a high energy testbeam~\cite{mathes-3D_2008}. (b) charge distribution for tracks perpendicularly entering the column electrode. (c) charge distribution for tracks entering the depleted region away from the electrode.}
\end{figure}

The obvious advantages of 3D - detector structures are fast and full charge collection at low bias voltages, as a consequence of which they are more radiation tolerant than comparable planar devices. Also the edge of the sensor itself can be an electrode guaranteeing a large active area. The fabrication of 3D sensors is a non-standard technology. Commercial vendors are currently investing in this new technology. 3D - pixel devices with electrodes reaching entirely through the substrate~\cite{3D-manufacturing,stanford-microlab} as shown in fig.~\ref{3D-pixels}(a) have been studied in a CERN SPS test beam~\cite{mathes-3D_2008}.
The devices show full depletion already at voltages of only 10~V. In order to study the effect of the vertical columns on the charge collection fig.~\ref{3D-testbeam}(a) shows the efficiency map of a pixel cell, scanned by the beam with a reference device resolution of about 5$\mu$m. While the charge distribution exhibits a nice Landau-shape in the depleted area (fig.~\ref{3D-testbeam}(c)), the collection is inefficient in the electrode itself (fig.~\ref{3D-testbeam}(b)) as expected. As the electrodes in the tested devices~\cite{stanford-microlab} reach completely through the bulk, the efficiency for track reconstruction for perpendicularly incident tracks is reduced to (95.6 $\pm$ 0.1)\%.
However, for inclined tracks, which is the typical case at colliders, the detection efficiency recovers to 99.9$\%$ at inclination angles of 15$^\circ$~\cite{mathes-3D_2008}. The spatial resolution obtained for perpendicular incidence is 12$\mu$m~\cite{mathes-3D_2008}.

\subsubsection*{Diamond pixel sensors}
Most pros and cons of diamond compared to silicon are due to its about four times larger band gap. As a consequence there is no thermal generation of charge carriers and no leakage current; diamond is very radiation tolerant. Table~\ref{diamond-Si} compares diamond and silicon material parameters. While the signal is smaller by a factor of 2.2, the noise in diamond benefits from the absence of leakage current (no shot noise) and the fact that the smaller dielectric constant results in a smaller input capacitance to the amplifier. Hence also the 1/f and thermal noise contributions are reduced.
\begin{table}[!htb]
\begin{center}
\label{diamond-Si}
\begin{tabular}[htb]{cccccc}
\hline
& silicon & diamond \\
\hline
band gap (eV) & 1.1 & 5.5 \\
av. energy for e/h (eV) & 3.61 & 13.1 \\
density (g/cm$^3$) & 2.33 & 3.52 \\
dielectric constant $\epsilon_r$ & 11.9 & 5.7 \\
radiation length X$_0$(cm) & 9.38 & 12.13 \\
av. energy loss $\langle \frac{dE}{dx} \rangle$ (MeV/cm) & 6.1 & 3.9 \\
signal (MPV) per 300 $\mu$m (e$^-$) & 23000 & 10300 \\
signal (MPV) per 0.1 X$_0$ (e$^-$) & 7150 & 4200 \\
\hline \\
\end{tabular}
\caption{Comparison of some material properties of diamond and silicon.}
\end{center}
\end{table}

In addition, the radiation length is somewhat larger for diamond. As a result, for typical LHC hybrid pixel chip parameters (input capacitance, amplifier speed and filter times, transistor g$_m$, leakage current), the S/N figure per 0.1$\%$ X$_0$ for small leakage currents is in the same order as for silicon. It can be even larger by almost a factor of two when typical leakage currents of the order of 100 nA are assumed for Si at the end of the LHC lifetime. Diamond is also an excellent thermal conductor, a property which can possibly be exploited for intelligent integrated cooling concepts of pixel vertex detectors.

\begin{figure}[thb]
\begin{center}
\subfigure[Single crystal pixel detector]{
    \includegraphics[width=0.35\textwidth]{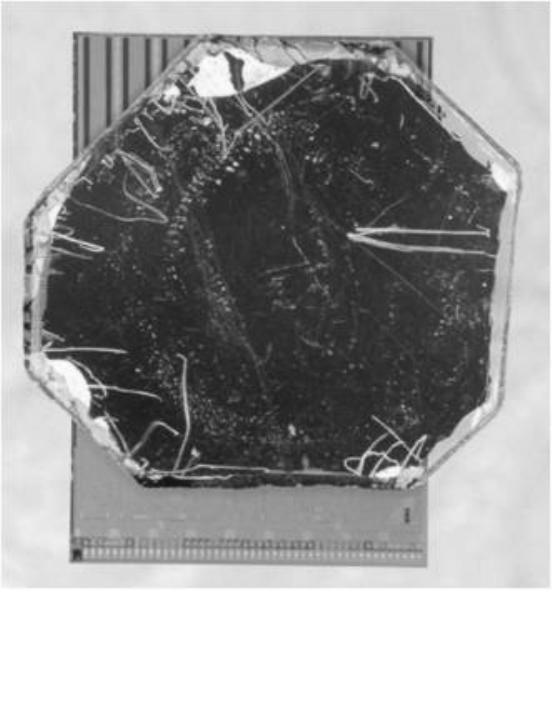}
    \label{diamond_1a}}
    \hskip 1cm
\subfigure[Charge distribution]{
    \includegraphics[width=0.35\textwidth]{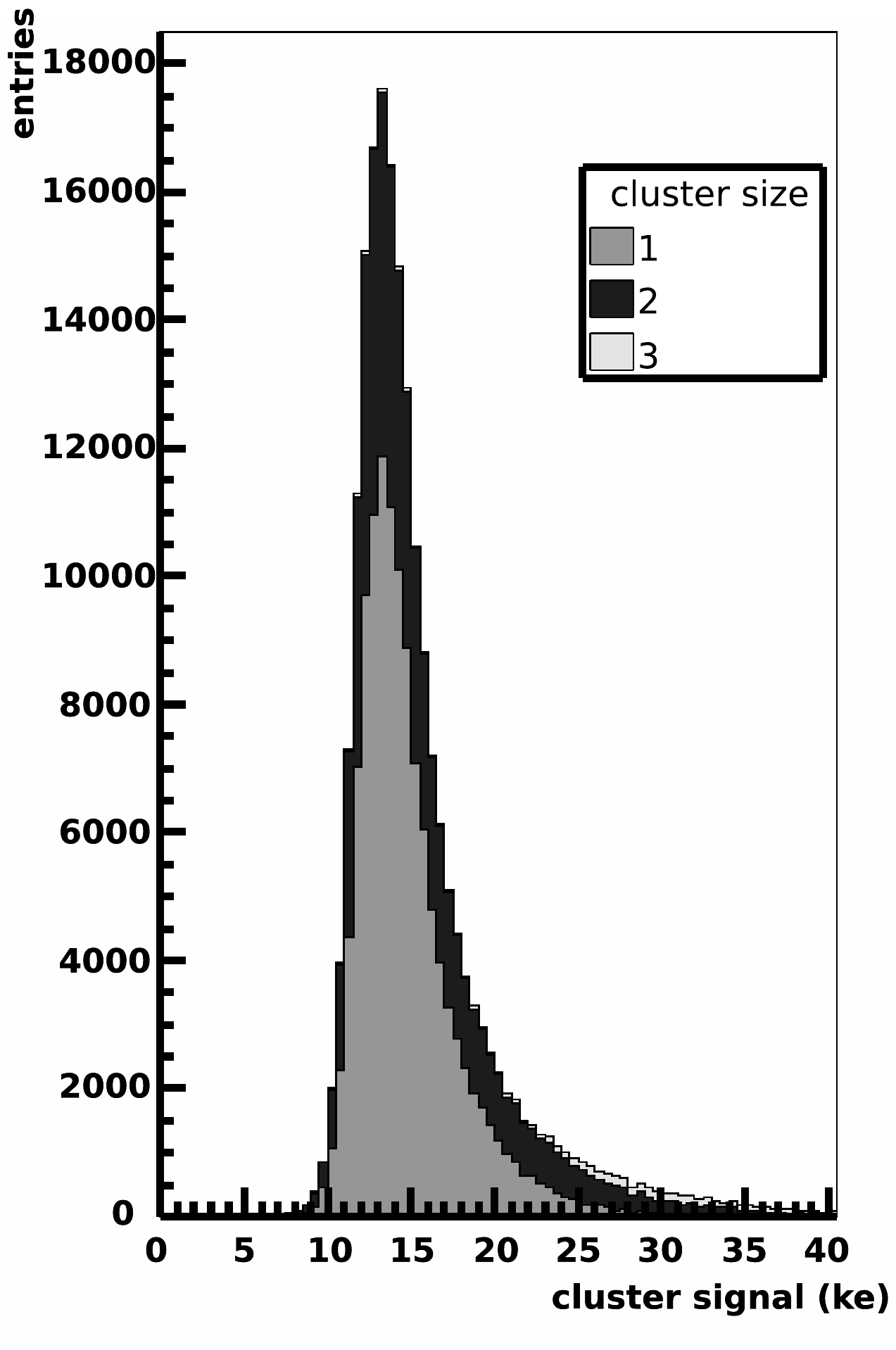}
    \label{diamond_1b}}
\end{center}
\caption[]{\label{diamond}(a) Single crystal diamond pixel detector bonded to ATLAS FE-I3 pixel chip~\cite{mathes-diamond_2008}. (b) Landau distribution measured with this detector using a 100 GeV pion test beam at 400 V bias voltage~\cite{mathes-diamond_2008}.}
\end{figure}

Single-crystal samples (scCVD) have been shown to follow the same damage curve as poly-crystalline (pCVD) material~\cite{RD42-LHCC-2008}. In~\cite{mathes-diamond_2008} test beam measurements using a single crystal pixel device mated to the ATLAS pixel FE-chip (fig.~\ref{diamond}(a)) report excellent charge collection properties leading to clean and narrow Landau distributions as shown in fig.~\ref{diamond}(b). The measured spatial resol\-ution is (8.9$\pm$0.1)\,$\mu$m for a 50\,$\mu$m pixel pitch. Grain structure effects, inherent to pCVD diamond, giving rise to horizontal polarization fields inside the diamond, have been shown to be absent in scCVD diamond sensors~\cite{mathes-diamond_2008}. Operation of a scCVD pixel device in the beam was possible with thresholds as low as 1300 e$^-$~\cite{mathes_priv}.

\subsubsection*{Planar silicon sensors}
While the aforementioned new sensor technologies may be a choice for the innermost, small area pixel layers at sLHC, it is difficult, simply for manufacturing and cost reasons, to imagine large area pixel detectors for the sLHC in these technologies. However, R\&D on the radiation hardness of planar silicon pixels has brought out, that sufficiently oxygen enriched silicon,
especially using the Magnetic Czochralski technique, is capable to sustain fluence levels larger than 10$^{15}$ cm$^{-2}$ with sufficient signal charge left. In
addition, p-type bulk material does not show type inversion. Hence, n$^+$ in p or n-bulk pixel sensors, for which the depletion zone forming diode remains at the pixel electrode side (or moves to the electrode side after type inversion for n-bulk) is the favorite option. In addition, the charge collection efficiency measurements do not show the so called \emph{reverse annealing} effect and signals close to 6000 e$^-$ at the end of a lifetime (10$^{16}$ n$_{eq}$ cm$^{-2}$) can be expected.

\begin{figure}[thb]
\begin{center}
\includegraphics[width=0.7\textwidth]{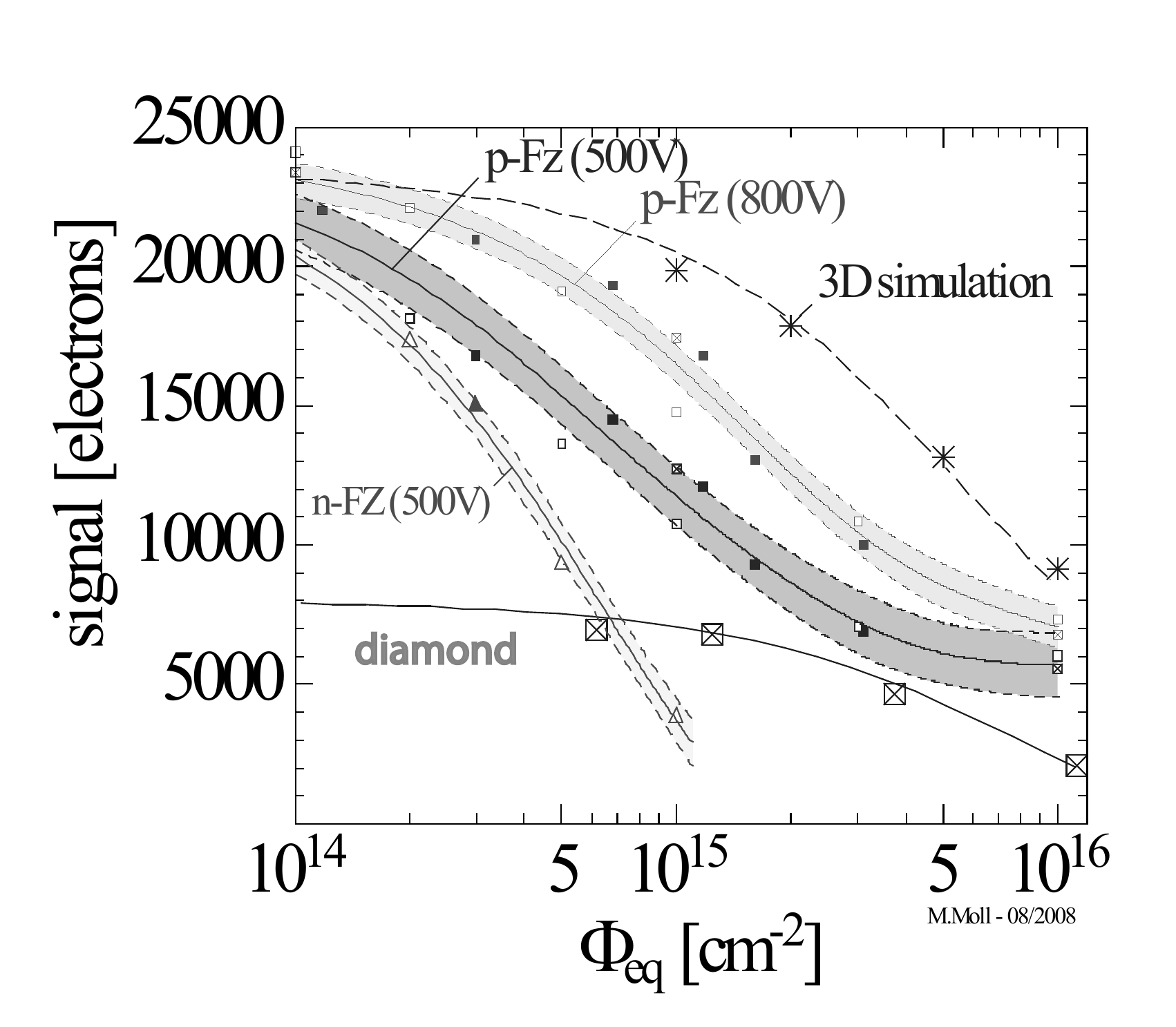}
\end{center}
\caption[]{\label{moll-comparison} Signal charge obtained after irradiation as a function of the radiation fluence for various planar silicon float zone (FZ) materials with n electrodes in p bulk (p-Fz) and p-electrodes in n bulk (n-Fz) as indicated by the bands (from~\cite{moll_priv}). Also included are simulations (Pennicard 2007) for 3D devices (stars) and measurements for diamond (crossed squares)~\cite{RD42-LHCC-2008}.
}
\end{figure}

Figure~\ref{moll-comparison} shows a summary of the current measurements~\cite{moll_priv,moll_PIX2005} of various planar silicon materials. Shown is the signal charge as a function of fluence~\cite{Casse_2008,moll_priv}, normalized to 1 MeV neutron equivalent damage (in Si). Also data for 3D silicon sensors (simulations only) and diamond\footnote{Note that an appropriate comparison would have to quote S/N for a given pixel configuration normalized to 1 MeV damage in diamond.}~\cite{RD42-LHCC-2008} are included in this plot.

To conclude this section on sensor materials, 3D and diamond sensors are attractive for the innermost pixel layers at the sLHC, but also planar Si sensors
may be suitable. Large area modules have so far only been built with pCVD diamond devices~\cite{kagan_pixel2005}. For scCVD diamond and 3D-silicon, wafer scale module production still must be demonstrated. For outer vertex detector layers the cost issue plays a major role. Here most likely planar silicon sensors are the only choice.

\subsubsection*{IC development for sLHC}
\begin{figure}[thb]
\begin{center}
\includegraphics[width=0.7\textwidth]{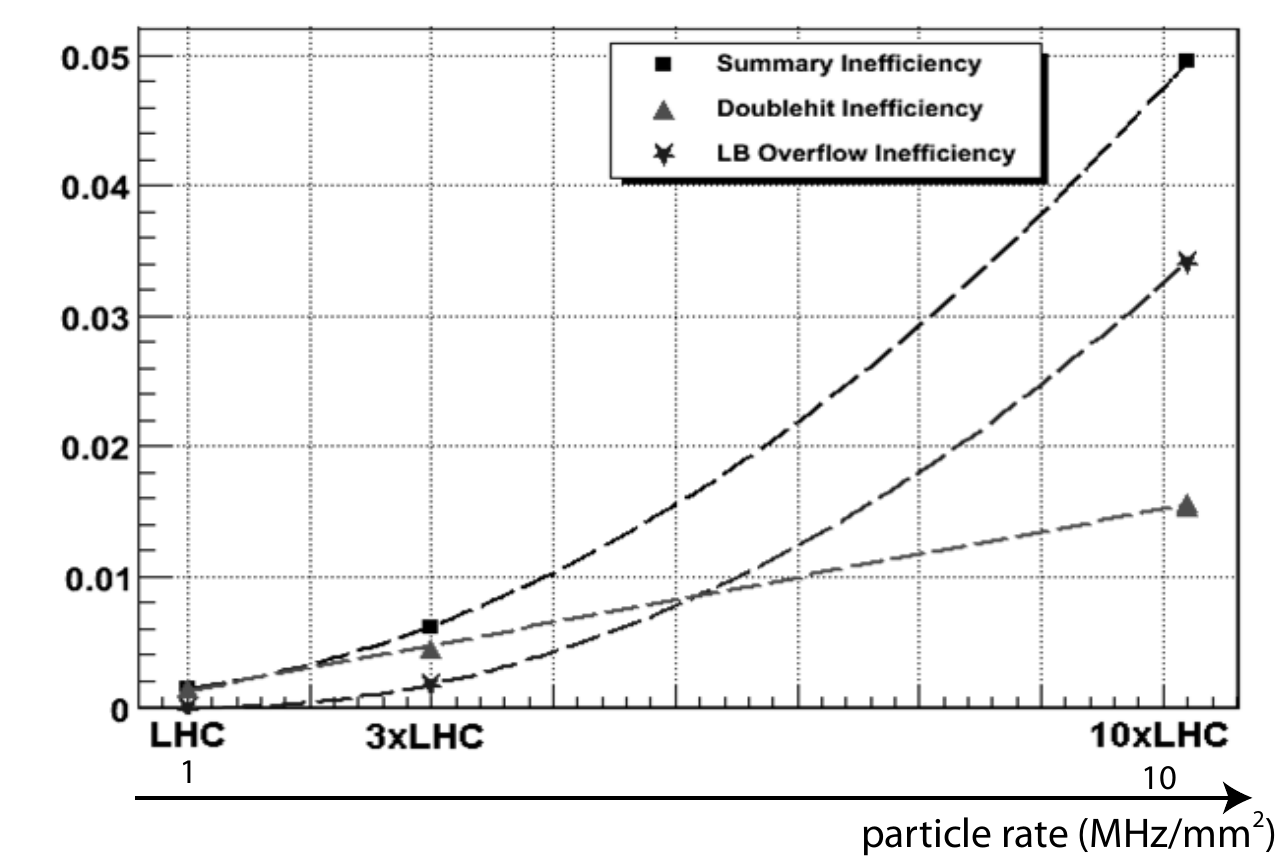}
\end{center}
\caption[]{\label{efficiency-simulation} Simulation of the hit detection inefficiency as a function of the particle rate, comparing LHC and superLHC~\cite{arutinov_priv}. Triangles denote buffer overflow inefficiency, stars denote double-hit inefficiencies in the pixel, squares show the total inefficiency.}
\end{figure}

The data rates expected at the sLHC ($>$ 10 MHz/mm$^2$) would -- with current chip architectures -- produce hit inefficiencies close to 100\%~\cite{arutinov_priv}. The bottleneck appears to lie in the fact that pixel hits are moved to the end of the pixel column within the R/O-chip where they ares stored until the trigger arrives. The obvious way out is to keep storage of pixel hits local in the pixel cell rather than moving them. A simulation~\cite{arutinov_priv} for the new architecture for ATLAS using physics data input shows in fig.~\ref{efficiency-simulation} the expected inefficiency as a function of the hit rate. The expected total hit inefficiency at sLHC rates is about 5\%~\cite{arutinov_priv}.
Another general trend is to increase the overall chip size as much as possible in a deep submicron technology (130 nm or 90 nm)~\cite{maurice_priv}. Experting that the yield does not scale inversely proportional to the chip area, the costs for flip-chipping will be much reduced. The total power is expected to be in the order of
200~mW/cm$^2$.

\section{Monolithic and semi-monolithic pixel detectors}\label{monolithic-pixels}
The two main streams of monolithic or semi-monolithic pixel developments are so-called DEPFET-pixels~\cite{kemmerandlutz88} and CMOS monolithic active pixels (MAPS)~\cite{meynants98_2,Turchetta2001}. Both have been developed for more than 10 years for an ILC vertex detector. They have now reached maturity levels such that real detectors are planned/built, for superBelle (DEPFET) and for the STAR detector at RHIC (MAPS), respectively.

\subsubsection*{DEPFET pixels}
\begin{figure}[thb]
\begin{center}
\subfigure[DEPFET double pixel cell structure]{
    \includegraphics[width=0.40\textwidth]{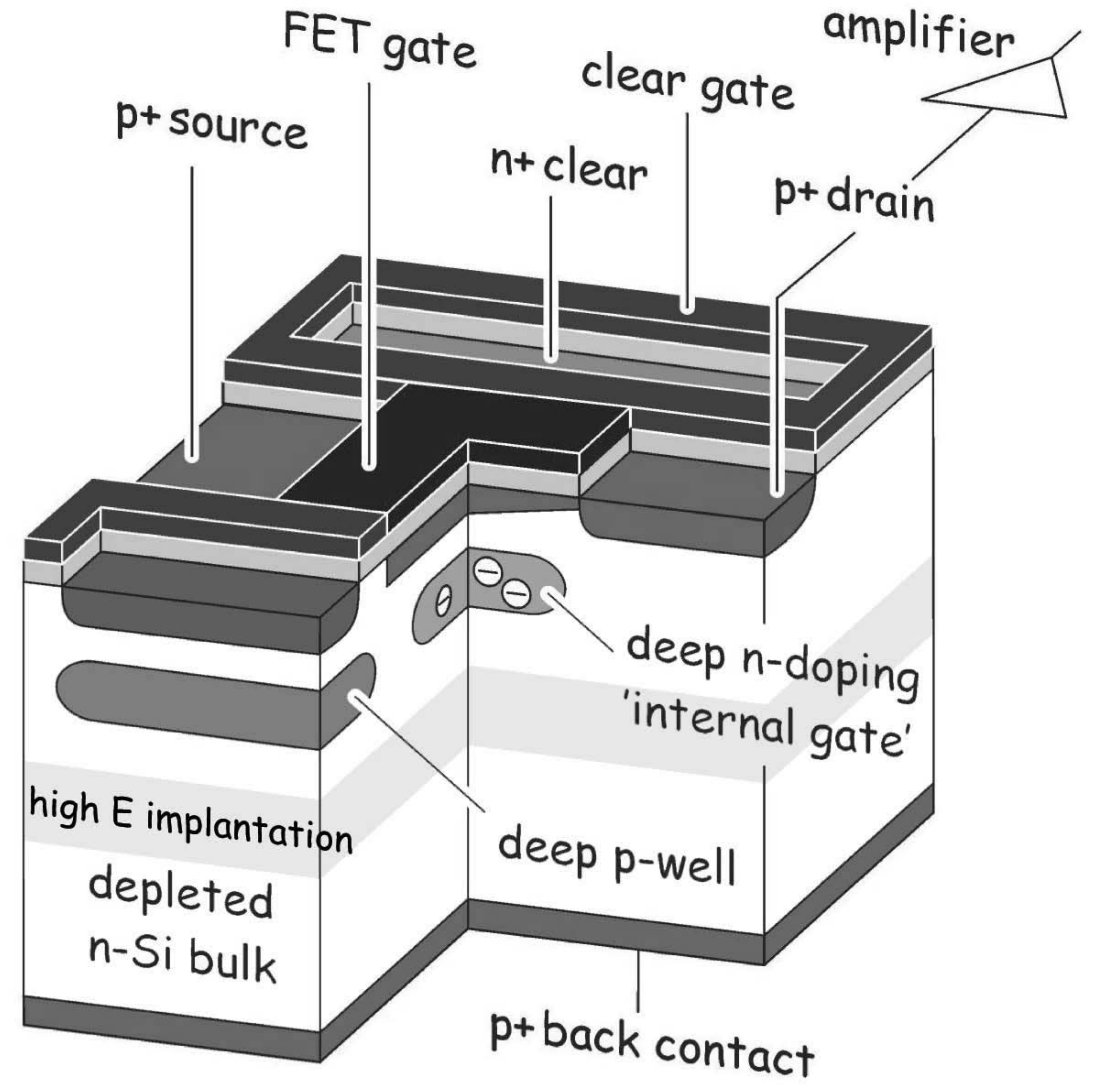}
    \label{DEPFET-1a}}
    \hskip 1cm
\subfigure[Symbolic electronic circuit for DEPFET operation]{
    \includegraphics[width=0.45\textwidth]{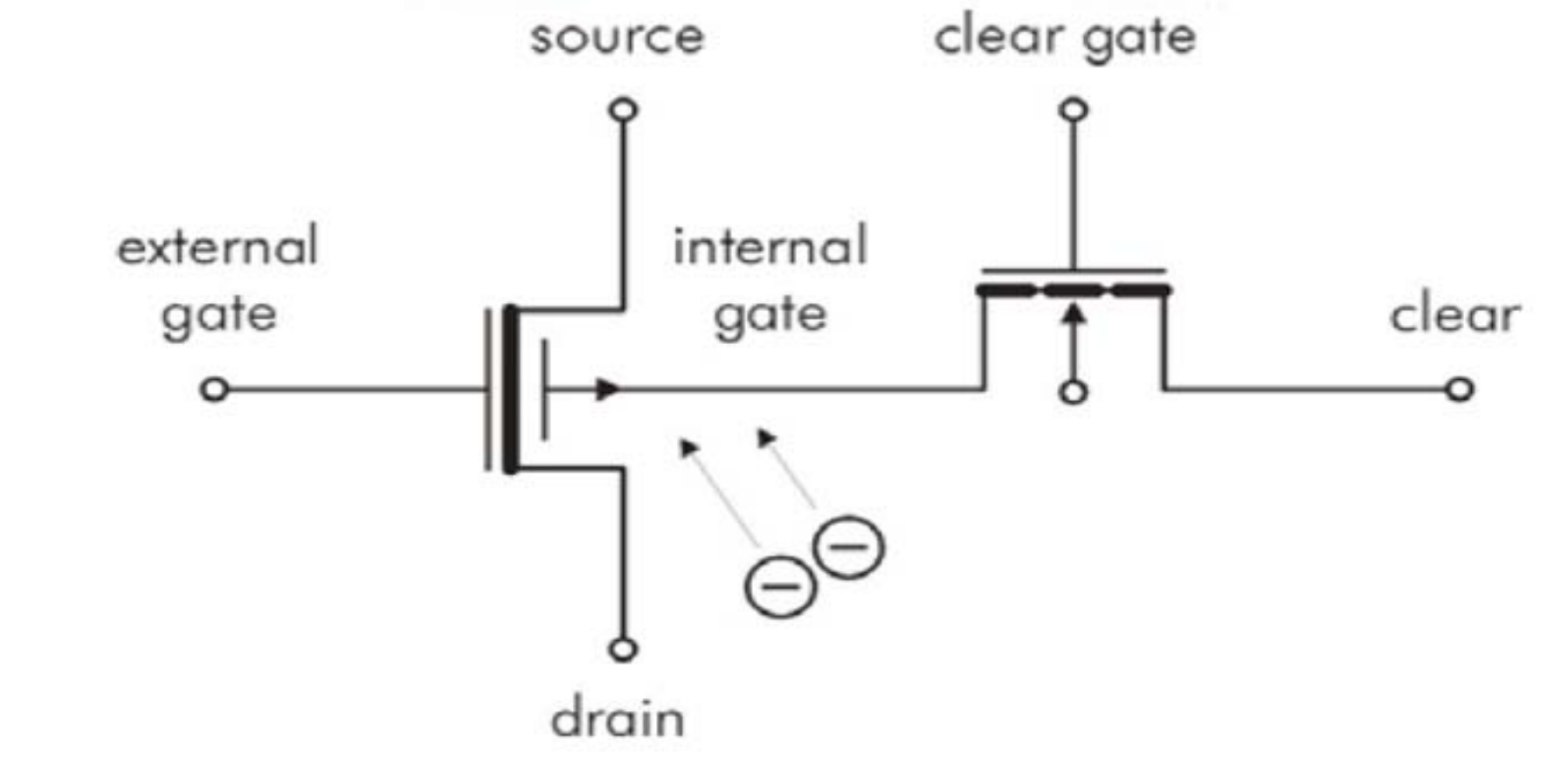}
    \label{DEPFET-1b}}
\end{center}
\caption[]{Principle of DEPFET pixels developed for ILC and superBelle~\cite{DEPFET-Pefi-07}.}
\end{figure}

{DEPFET pixels}~\cite{kemmerandlutz88,laci04,kohrs_pixel2005} are not truly monolithic devices as only the amplifying transistor is implemented in every pixel cell (fig.~\ref{DEPFET-1a}). Electrons generated in the fully (sidewards) depleted bulk are collected in a local potential minimum underneath the transistor channel (internal gate) and hence directly steer the channel current producing an output current signal at the drain. The device is controlled by four terminals: source, drain, internal and external gate (cf.~fig.~\ref{DEPFET-1b}). A full matrix is frame-readout by selecting a row of gate connected DEPFET-transistors using the external gate line, and the transistor currents of the selected row are read out column-wise at their drains~\cite{kohrs_pixel2005}. After readout the row is cleared by a dedicated clear pulse, which is steered by a common, but adjustable clear-gate voltage.
\begin{figure}[bh!]
\begin{center}
    \includegraphics[width=0.95\textwidth]{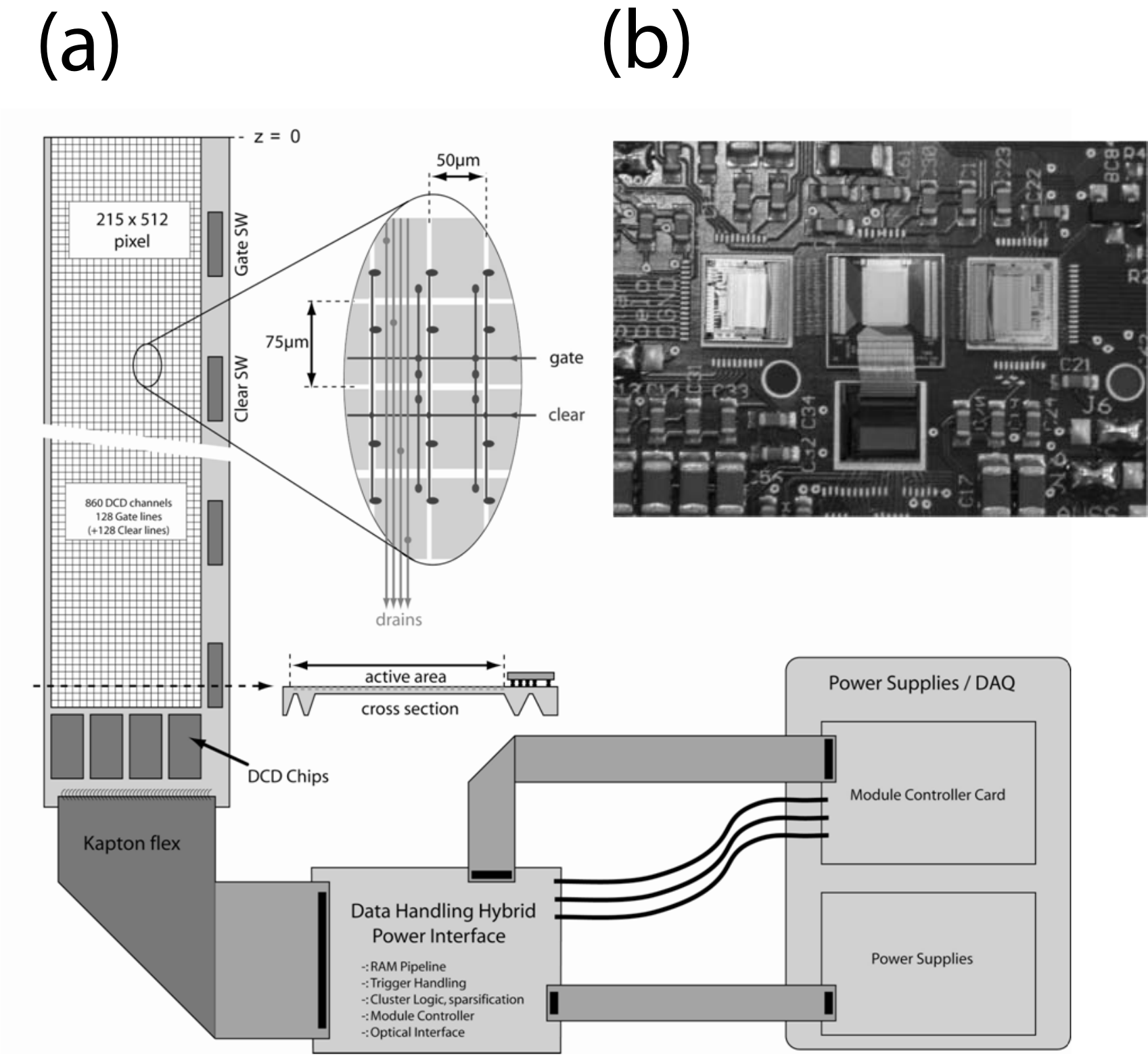}
\end{center}
\caption[]{\label{DEPFET-readout} (a) DEPFET sensor/IC arrangement and readout for the superBelle vertex detector. (b) Photograph of a DEPFET pixel hybrid with DEPFET sensor matrix (center), steering (left), clearing (right), and readout (bottom) ICs~\cite{DEPFET-Pefi-07}.}
\end{figure}
The advantages of the DEPFET principle are (a) a large signal due to charge collection in the fully depleted bulk, and (b) a low noise due to the small capacitance
(between internal gate and transistor channel), i.e. a very comfortable S/N ratio. For long shaping times noise figures below 2e$^-$ have been measured while for
high rate, high speed readout for ILC or superBelle operation the noise figure is typically 200 e$^-$ or above, leading to S/N values of 120 measured in test beams with 450 $\mu$m thick sensors. Using 50 $\mu$m thin sensors for superBELLE the S/N is expected to be between 20 and 40, while the radiation length of a detector layer will lie below 0.15\% X$_0$. The power consumption is 60 mW/cm$^2$ due to the selective row R/O which allows all other rows to be switched off during data dating. The position resolution is below 2$\mu$m~\cite{velthuis2007}.
The disadvantage compared to more monolithic approaches is the need for a steering IC and a current processing IC which are external to the sensing pixel matrix (see photo in fig.~\ref{DEPFET-readout}(b)), i.e. CMOS processing is done externally. For superBelle four pixels (50$\times$75 $\mu$m$^2$) are read out in parallel to both sides to gain an 8-fold frame rate increase (see fig.\ref{DEPFET-readout}).
The radiation tolerance of DEPFETs has been tested to the following fluences/doses: 0.9 Mrad $\gamma$, 3 x 10$^{12}$ p/cm$^2$, 2 x 10$^{11}$ n/cm$^2$. Noise due to leakage current increase of 20 -- 95e$^-$ and threshold shifts of $\sim$4V have been observed. Since there is only one transistor in every pixel, this shift is still comfortably compensated by an appropriate tuning of the voltage of the external gate, delivered by the steering chip.

\subsubsection*{MAPS-epi}
\begin{figure}[thb]
\begin{center}
\subfigure[MAPS-epi principle]{
    \includegraphics[width=0.45\textwidth]{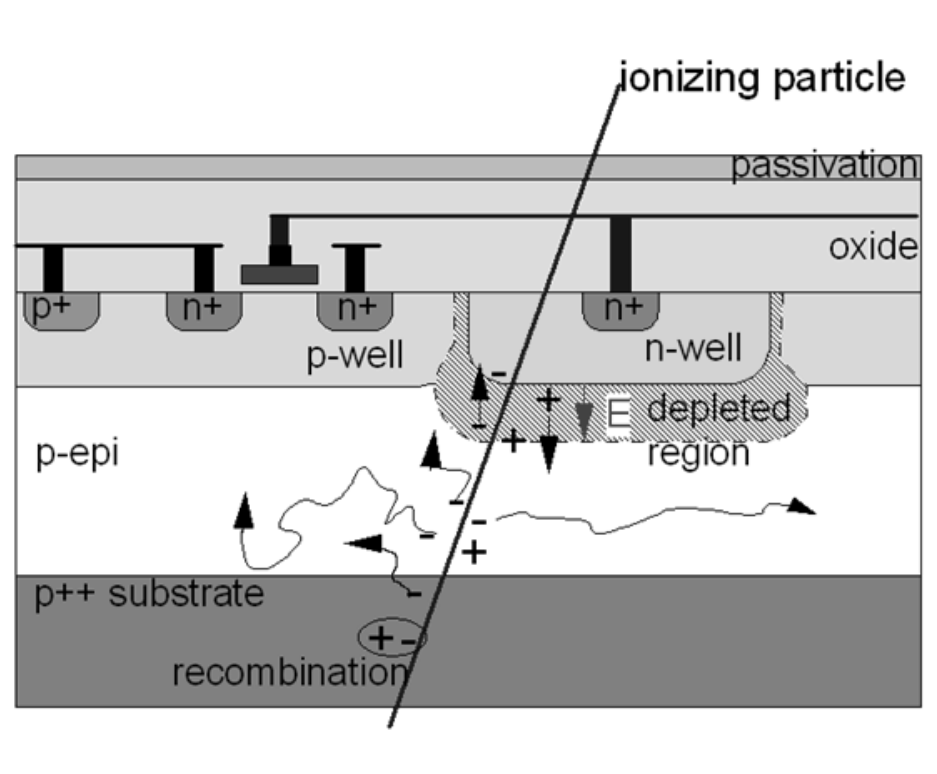}
    \label{MAPS-epi-principle}}
    \hskip 0.1cm
\subfigure[MAPS-epi structure with deep p-well]{
    \includegraphics[width=0.48\textwidth]{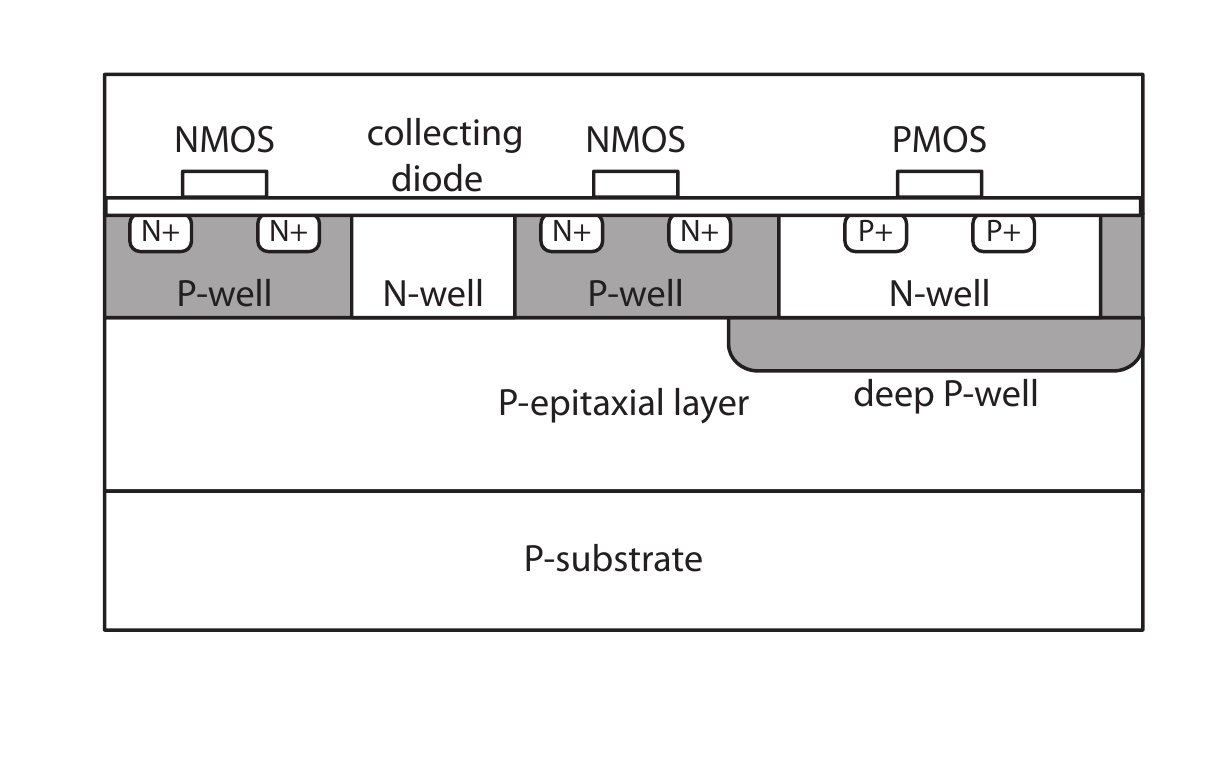}
    \label{deep-pwell}}
\end{center}
\caption[]{\label{MAPS} (a) Principle of a MAPS-epi detector~\cite{MSzleniak_2008}. (b) MAPS-epi pixel structure with a deep p-well to shield the PMOS transistors~\cite{turchetta_deeppwell}.}
\end{figure}

{MAPS-epi} pixels~\cite{meynants98_2,Turchetta2001} exploit the fact that in many CMOS technologies, especially those for CMOS camera applications, there is a (10--15~$\mu$m) thick epitaxial layer between the silicon substrate layer and the CMOS electronics layer. Electrons generated in this layer thermally diffuse towards an n-well collecting diode which is connected to the gate of a transistor. The fill factor for particle detection is 100\%. The signal of a minimum ionizing particle is typically diffused over several pixels and amounts in total to about 1000 e$^-$ or less. Low noise readout is therefore the challenge in this development.
MAPS-epi structures as sketched in fig.~\ref{MAPS-epi-principle} are is the currently most matured development towards monolithic pixel detectors. As the collection diode is an n-well, no competing n-wells are possible in the active pixel area and hence no PMOS transistors. CMOS signal processing must thus be deferred to the boundaries of the pixel chip.
The main advantages of MAPS-epi devices in comparison to hybrid pixels are the small possible pixel sizes and the small possible material budget. Radiation tolerance and readout speed, however, are challenges for MAPS-epi pixels.

For operation of a MAPS-epi pixel matrix with self biased structures a minimum of 2 NMOS transistors (source follower, select) is needed~\cite{MAPS-STAR-Dulinksi}. The signal is obtained by subtracting two frame readouts (correlated double sampling, CDS) and subsequent correction for baseline shifts, 1/f noise and fixed pattern noise. Operation of MAPS-epi matrices in the EUDET telescope~\cite{EUDET-1} in CERN and DESY test beams yields typical figures of S/N $\sim$ 11 and resolutions of $\sim$ 3.3$\mu$m. For the STAR vertex detector at RHIC~\cite{STAR-MAPS-Greiner,STAR-MAPS-Xu_2006} prototype developments with pixel arrays up to 320$\times$640 pixels and 30$\mu$m pixel pitch integrate signal processing CMOS circuitry outside the active pixel area. The next prototype generation (Mimosa 8/16), ready for fabrication, features
apart from analog R/O, fast digital R/O with on-chip discriminators and zero suppression, as well as in-pixel CDS~\cite{MSzleniak_2008}.

\begin{figure}[thb]
\begin{center}
\includegraphics[width=0.45\textwidth]{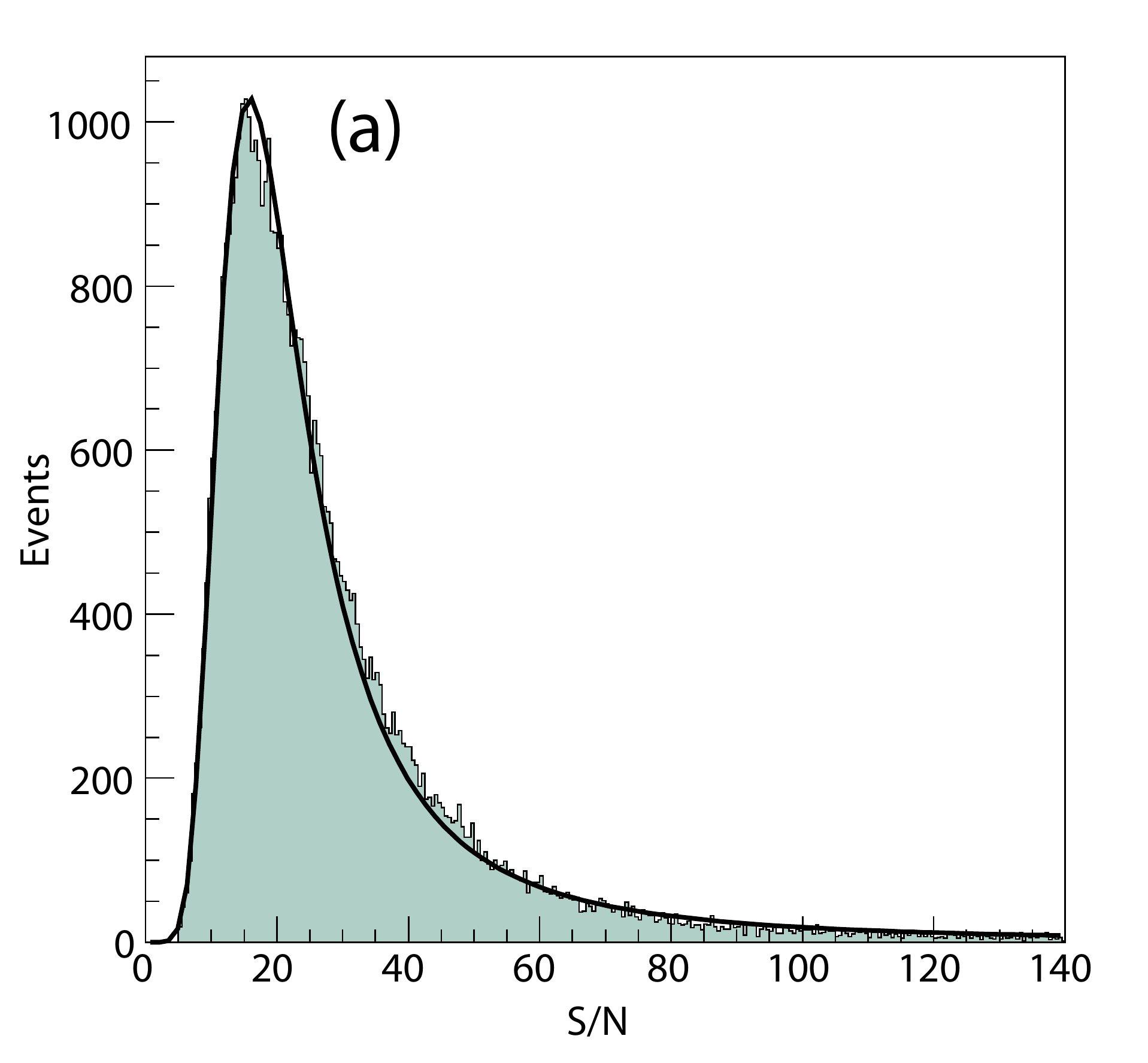}
\includegraphics[width=0.45\textwidth]{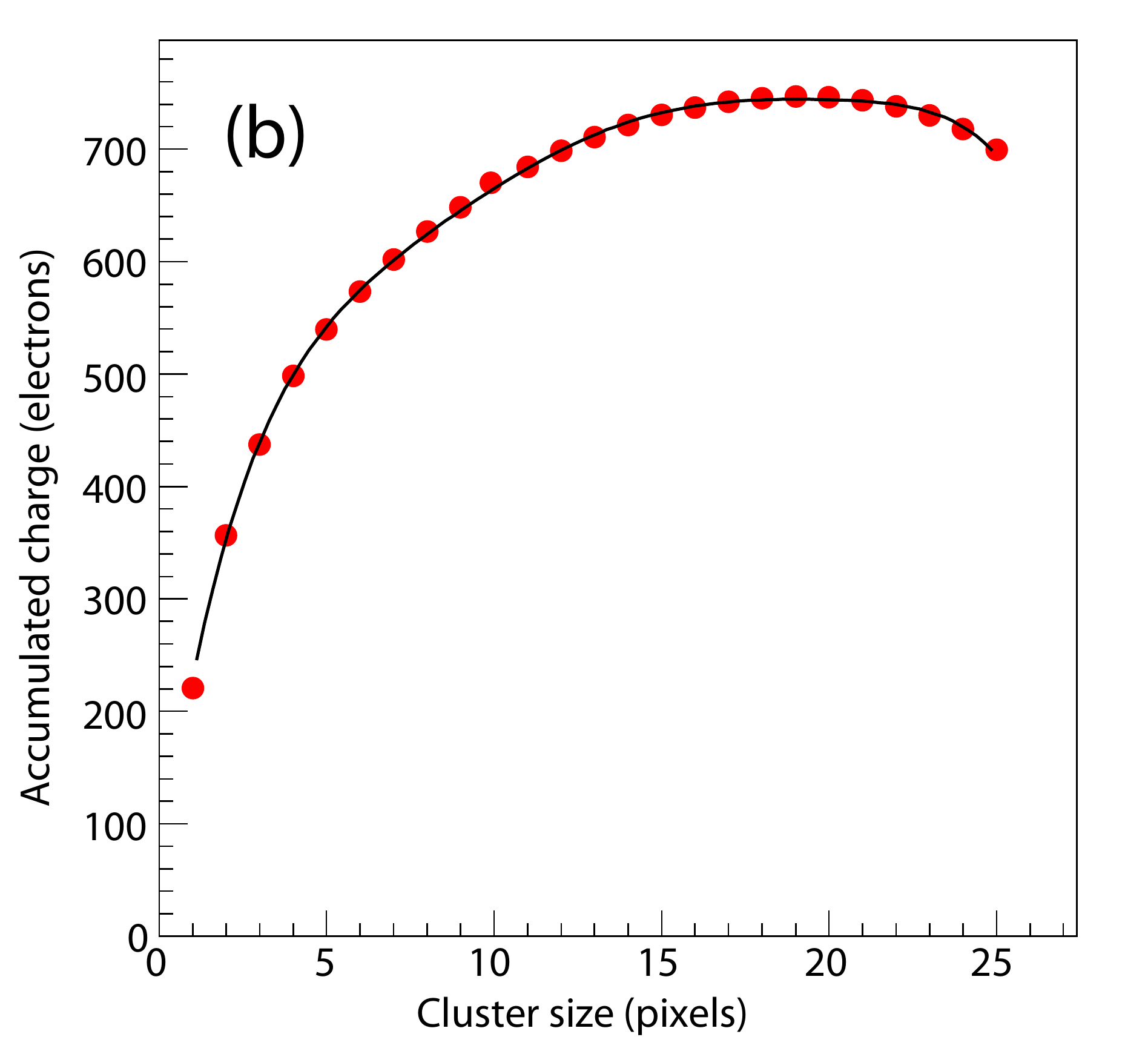}
\end{center}
\caption[]{\label{MAPS-STAR} (a ) Signal-to-noise distribution for the seed pixel measured with a \\ MAPS-epi prototype matrix for STAR@RHIC, (b) signal size as a function of the cluster size (from~\cite{MSzleniak_2008}).}
\end{figure}

Some typical results obtained with the AMS 0.35 $\mu$m technology~\cite{Degerli2006}, still read out with an integration time in the ms range, are shown in fig.~\ref{MAPS-STAR}, showing the charge distribution of the seed pixel expressed as S/N with a most probable value at 16.6, and the total cluster charge as a function of cluster size (fig.~\ref{MAPS-STAR}(b)) which reaches an optimum for about 20 summed pixels of 760e$^-$. The spatial resolution expected for the new  prototypes (Mimosa 16) is about 5 $\mu$m at a fake hit level of 10$^{-5}$.
For the STAR detector, ladders with 10 MAPS chips of 2x2 cm$^2$ are foreseen. The chips will be thinned to 50 $\mu$m allowing for the design goal of 0.28\% X$_0$ per ladder. The expected power budget is 100 mW/cm$^2$ and the integration time goal is 200 $\mu$s. The devices must sustain a radiation dose of 300 krad/yr and a fluence of $\sim$ 1$\times$10$^{13}$/cm$^2$ n$_{eq}$/yr. Depending on the ultimate radiation hardness of these sensors, it might be required to replace the detector yearly.

\begin{figure}[thb]
\begin{center}
\subfigure[]{
    \includegraphics[width=0.55\textwidth]{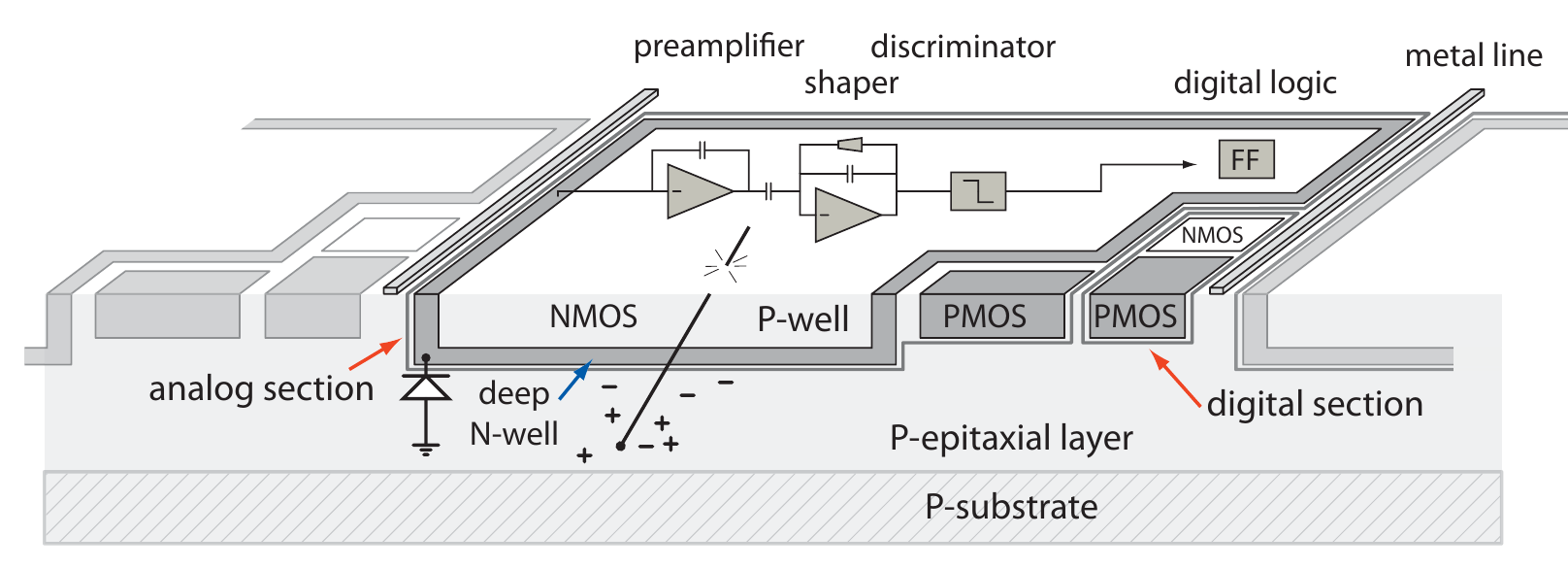}
    \label{deep-nwell-a}}
    \hskip 0.2cm
\subfigure[Cluster charge distribution]{
    \includegraphics[width=0.38\textwidth]{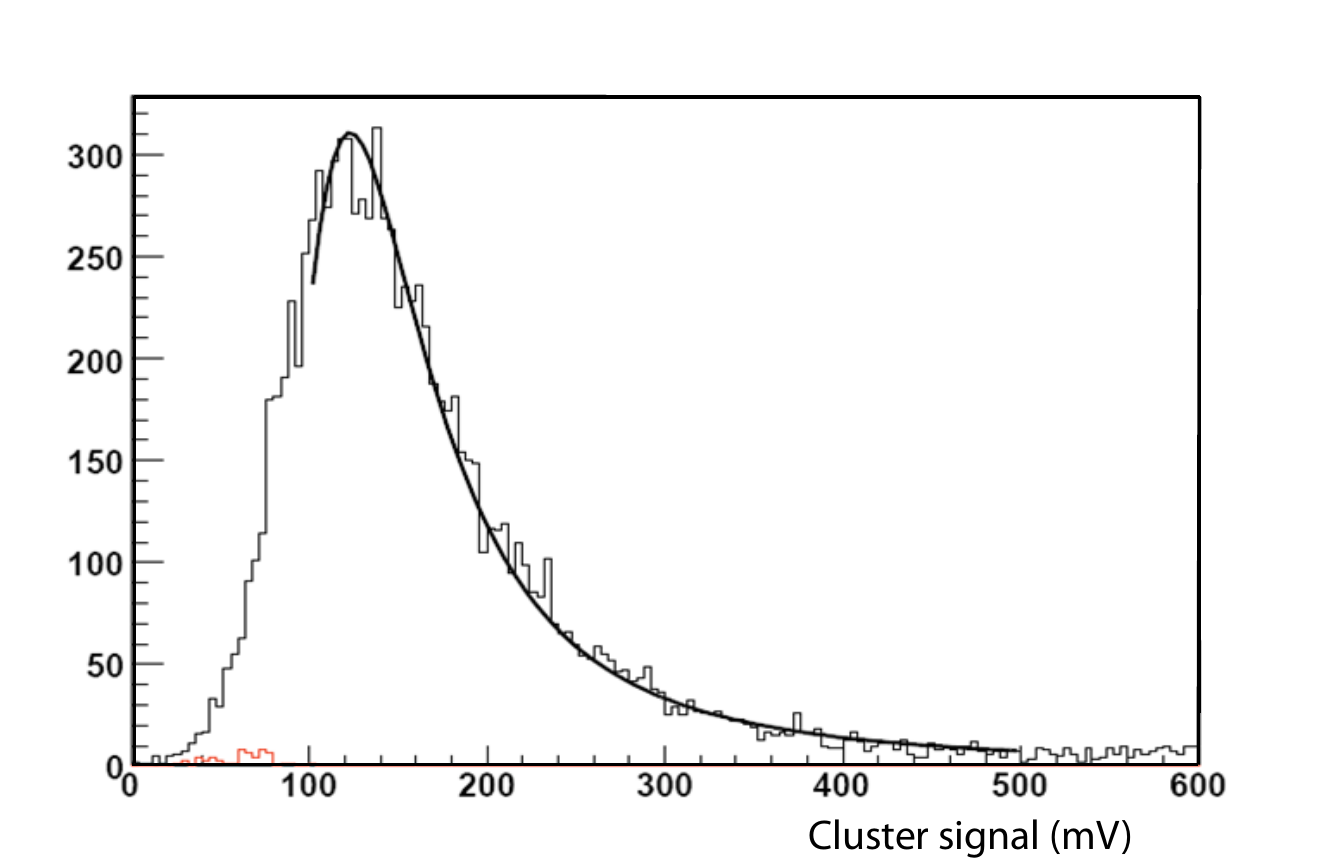}
    \label{deep-nwell-b}}
\end{center}
\caption[]{\label{deep-nwell}  (a) MAPS-epi development using a deep n-well structure (schematic layout~\cite{deep-nwell2007}). (b) measured cluster signal (mV) with the APSEL3 prototype chip in response to a $^{90}$Sr source~\cite{rizzo_2007}. The S/N is 23.}
\end{figure}

Further R\&D regarding MAPS-epi devices focuses on an improved charge collection and on full CMOS circuitry also in the active area (i.e. PMOS transistors).
One very promising approach~\cite{deep-nwell2007} uses a \emph{deep n-well} which extends over 90\% of the active area (fig.~\ref{deep-nwell}(a)). The deep n-well provides a higher single pixel charge collection efficiency and at the same time protects against to large charge losses to competing n-wells housing PMOS transistors.
Thus full CMOS circuitry is possible such that complete signal processing chains like charge sensitive preamplifier, discriminator, shaper plus logic can be realized within the pixel cell. The response of a prototype matrix (APSEL3~\cite{rizzo_2007}) to a $^{90}$Sr $\beta$-source is shown in fig.~\ref{deep-nwell}(b). The S/N = 23 is measured. Pixel matrices (32 $\times$ 128 pixels) with 50 $\mu$m pitch have been tested which collect the charge generated by an $^{55}$Fe 6 keV X-ray source in one pixel~\cite{rizzo_2007}.

Another approach uses a \emph{deep p-well}~\cite{turchetta_deeppwell} in a quadrupel well process called INMAPS to shield the n-wells containing PMOS transistors. This is shown in fig.~\ref{deep-pwell}. The electrons are pushed away from the p-well thus shielding the n-well while the charge collecting n-well diode is not shielded. Hence PMOS transistors can be used in the active area together with NMOS. Several circuit designs with full signal processing using more than 150 transistors have been submitted and tested~\cite{turchetta_deeppwell}. The first results show an increase in charge collection efficiency at the n-well-diode by more than a factor of two~\cite{turchetta_deeppwell}.

\subsubsection*{MAPS-SoI}
\begin{figure}[thb]
\begin{center}
\subfigure[MAPS-SoI structure]{
    \includegraphics[width=0.4\textwidth]{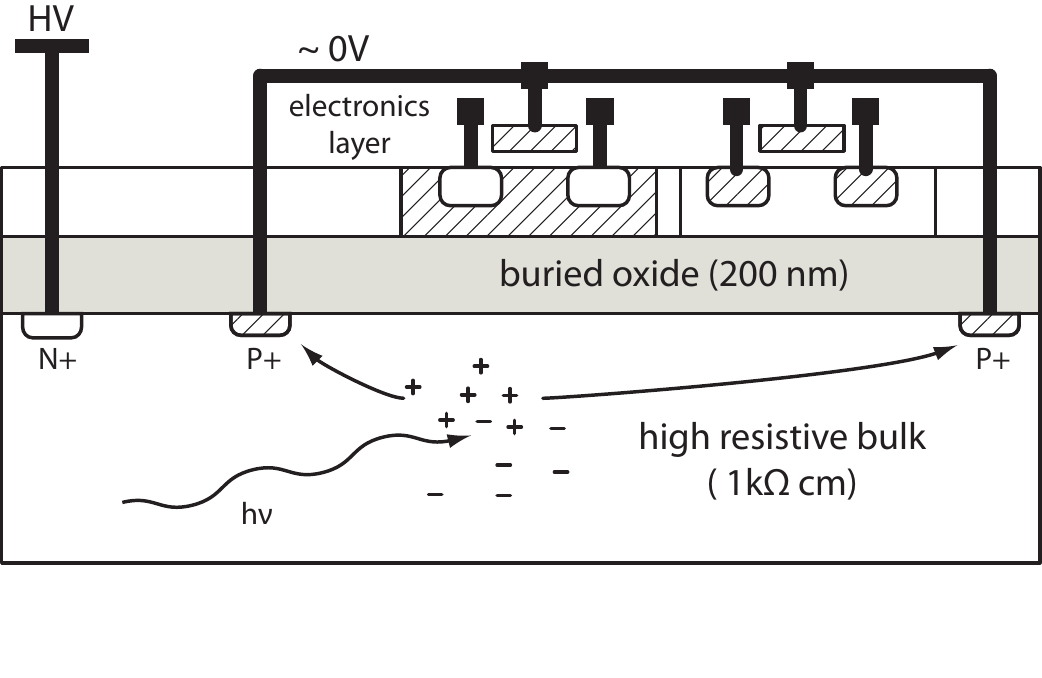}
    \label{MAPS-SoI}}
    \hskip 0.3cm
\subfigure[backgate effect]{
    \includegraphics[width=0.53\textwidth]{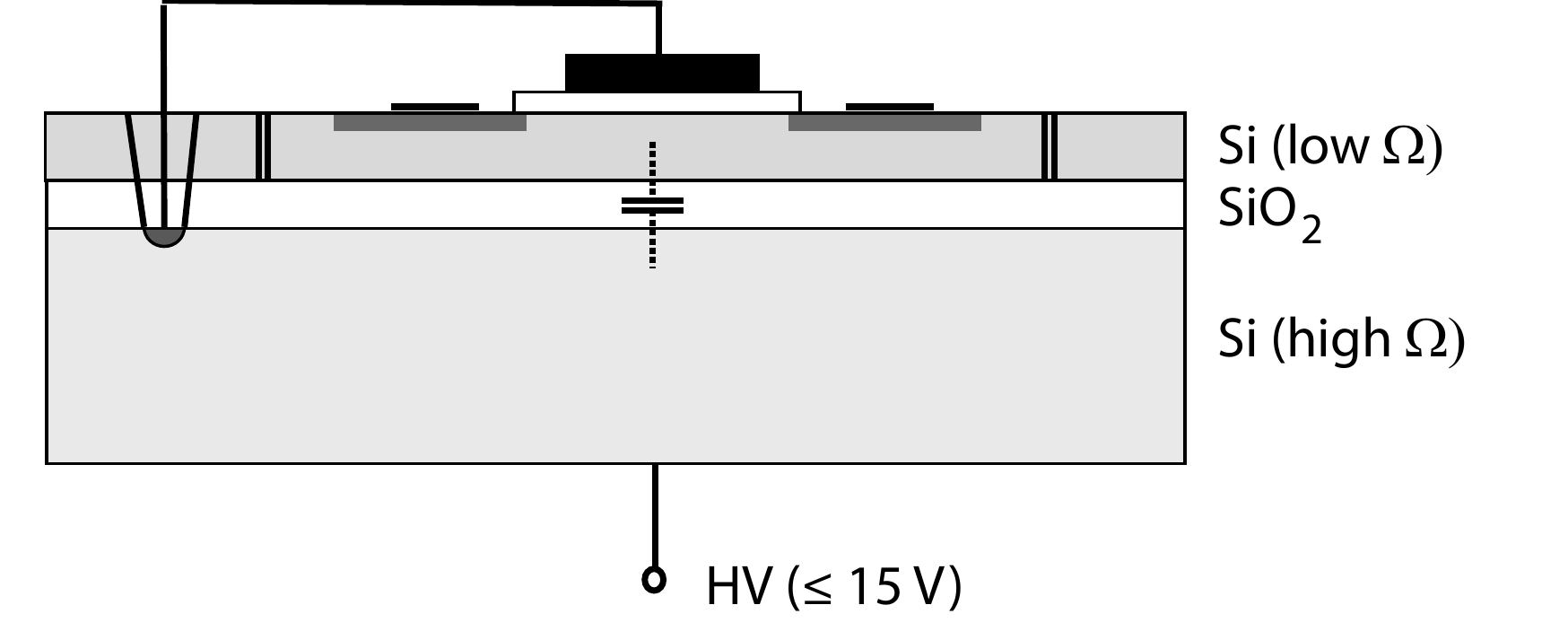}
    \label{backgate_effect}}
\end{center}
\caption[]{(a) Principle of a MAPS-SoI structure. (b) Illustration of the backgate-effect in monolithic active pixels in SoI technology (simplified).}
\end{figure}

The silicon on insulator (SoI) technique is an established technique for transistor isolation. The thin ($\sim$100 nm) Si CMOS layer is isolated by a SiO$_2$ layer and by shallow trenches; there is no connection to the bulk, every transistor is its own island. MAPS-SoI (fig.~\ref{MAPS-SoI}) exploits the SoI technique by combining it with the MAPS technique. For the handle wafer a fully depleted bulk is used to which a contact with the CMOS layer is made by a via contact connecting the signal electrode to the gate of the amplifying transistor. This development comes close to the dream of a monolithic pixel device: large signal charge collection in a high ohmic bulk combined with full CMOS signal processing. First developments in this direction were already made some years ago~\cite{fpengg1995,HAPS-SOI} with limited success due to non-industrial manufacturing. A new initiative has recently been started at Fermilab triggered by the availability of the 150~nm and 200~nm processes at OKI (Japan) which allows to interrupt the commercial CMOS process chain for via-processing. The development is still in the R\&D-phase. In terms of granularity and material budget MAPS-SoI have the same advantages as MAPS-epi detectors. The S/N figures of MAPS-epi pixels, however, are greatly improved. Probably the main problem of MAPS-SoI currently is the so-called backgate effect, illustrated in fig.~\ref{backgate_effect}. The high-ohmic substrate couples into the transistor channel and acts as a second gate. Threshold shifts of more than 100 mV are the consequence, not tolerable for CMOS circuits. The effect depends on the size of the bias voltage and is less pronounced when the substrate is only partially depleted with lower biasing voltage. Ideas to overcome this problem, like the use of a thicker buried oxide layer or a dense matrix of p$^+$ implants to lower the potential at the interface surface, exist and are under investigation~\cite{deptuch-priv}.
%

\section{3D integration}
\begin{figure}[tb]
\begin{center}
\subfigure[3D integration]{
    \includegraphics[width=0.45\textwidth]{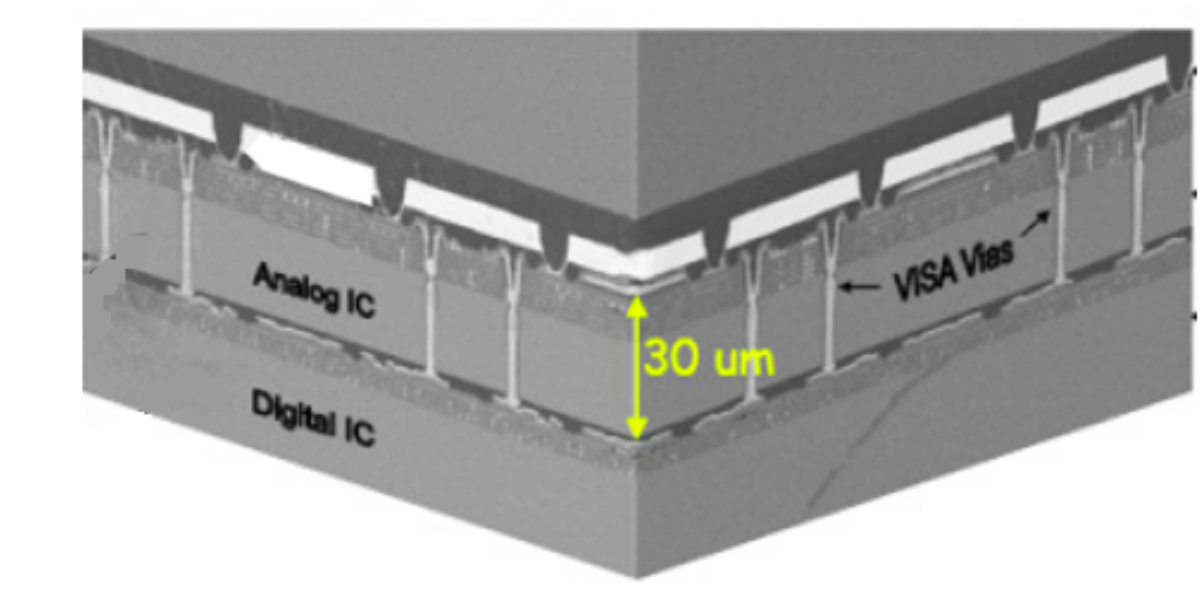}}
    \hskip 0.5cm
\subfigure[Fermilab 3D design]{
    \includegraphics[width=0.45\textwidth]{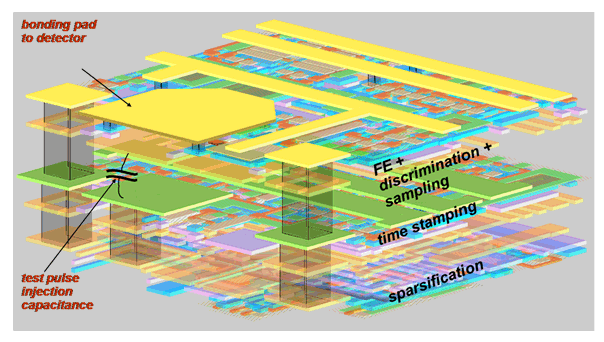}}
\end{center}
\caption[]{(a) Sketch of the 3D-integration technique, integrating several electronic layers~\cite{yarema_twepp2008}, (b) Fermilab 3D design with 3 tiers for different electronic tasks.\label{3D-integration}}
\end{figure}
3D-integration is a very important trend for compact IC-building with an enormous industrial market. This trend is extremely attractive also for pixel detector developments. Fig.~\ref{3D-integration}(a) illustrates the principle: one multilayered block of several thin layers (tiers) for different electronics tasks vertically connected by vias. This allows, for example, to separate analog and digital circuitry and to design as a top layer a dedicated opto-interface. The various layers could in principle even be made in different technologies. There are two approaches. In the one called \emph{vias first} the vias are embedded in the circuit at the foundry before or after the transistors are formed. In the \emph{vias last} process, the vias are added to the wafers after the CMOS processing has been completed, often done at a third party vendor. A sensor layer can thus be included for examples as the first tier in CMOS processing chain as in the case for CMOS active pixel sensors or can be bonded to the tier stack after processing as in the case of fully depleted sensors. More details can be found in~\cite{3D-handbook} and in~\cite{yarema_twepp2008,deptuch_vertex2008,yarema_vertex2007}.
3D-integration is driven by industrial goals, since it promises reduced resistances, capacitances, and hence reduced interconnect power and cross talk, as well as reduced size. The particle physics community in the US and Europe has become interested 2-3 years ago. Fermilab has made 3D integration attempts~\cite{yarema_LHCEE2006,deptuch_vertex2008} with different foundries and have among others submitted designs
including three different tiers for (1) front-end amplification, discrimination and sampling, (2) time stamping, and (3) sparsification as is shown in  fig.~\ref{3D-integration}(b). The sensor is not yet included.

Another 3D bonding R\&D (vias last) is followed by MPI-Munich together with Fraunhofer IZM~\cite{SLID}. The so-called Solid-Liquid-Interdiffusion (SLID) process is a technology to process and bond vias of different tiers by interconnecting them using tungsten plugs.

\section{Conclusions}
\begin{figure}[h!!!]
\begin{center}
\includegraphics[width=0.9\textwidth]{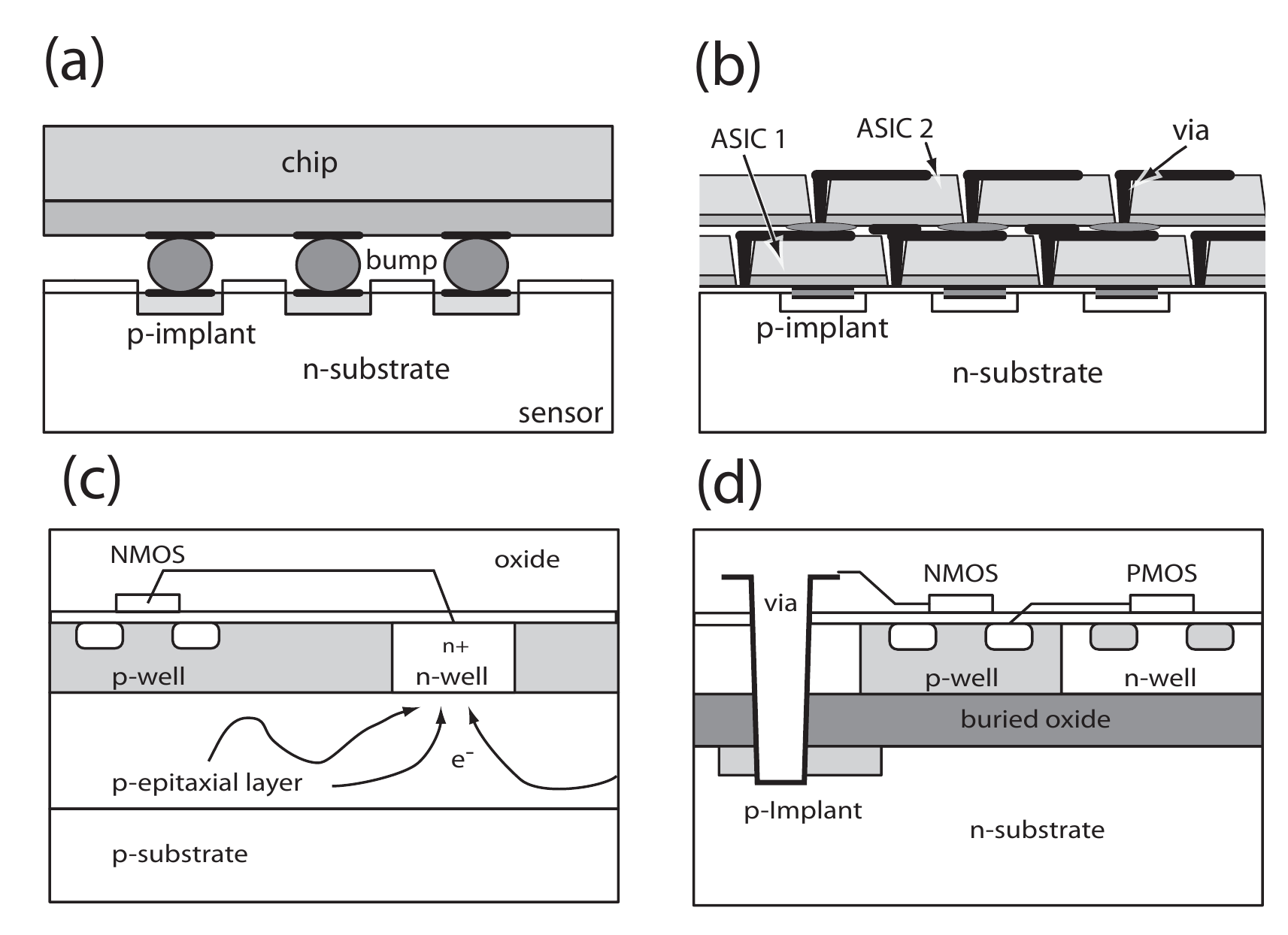}
\end{center}
\caption[]{\label{summary} Classification of pixel developments: rows denote hybrid technologies (a),(b) and monolithic technologies (c)(d);  columns denote standard (a),(c) versus advanced (b),(d) technologies. (a) hybrid pixels, (b) 3D integration (hybrid or monolithic), (c) MAPS-epi, (d) MAPS-SoI.}
\end{figure}

Figure~\ref{summary} summarizes the attempted classification of this paper: hybrid versus monolithic pixels and present versus advanced technologies. 3D-integration
can in fact be assigned to both, hybrid and monolithic approaches.
Hybrid pixels have reached the necessary maturity level for large detectors with the construction of the LHC experiments. This technology is
the only choice for sLHC detectors due to the large particle rates and the radiation environment. Heavy R\&D is necessary on new Si sensor and other materials, low mass module development and IC design capable of high rates. Also 3D-integration is a likely technology used for advanced hybrid pixels.
On the monolithic side DEPFET pixels and MAPS-epi have now reached a maturity level such that real vertex detectors are being built (superBELLE and STAR, respectively). The ultimate monolithic detector with fully depleted sensor and several layers of electronics circuitry has entered a new phase of R\&D with commercial vendors. 3D integration enters here in a natural way.

\section*{Acknowledgments}
The author would like to thank
Michal Szelezniak, Grzegorz Deptuch, Laci Andricek, Michael Moll, Woitek Dulinski,
Renato Turchetta, Hans-G\"unther Moser, Hans Kr\"uger, Fabian H\"ugging, Marlon Barbero, David Arutinov, Giovanni Darbo,
Cinzia da Via$^\prime$, Valerio Re, and Harris Kagan for providing information and for useful discussions. He thanks Manuel Koch for help with the figures.
%
\bibliographystyle{elsart-num-sort}
\bibliography{Wermes_PSD8_081121}


\end{document}